\documentclass[twocolumn,showpacs,preprintnumbers,amsmath,amssymb]{revtex4}

\usepackage{epsfig}
\usepackage{dcolumn}
\usepackage{bm}
\usepackage{float}

\begin{document}

\title{\boldmath $a_0^+(980)$--resonance production in
  $pp\rightarrow d K^+\bar{K^0}$ reactions close to threshold}
    
\author{
  V.~Kleber,$^1$\footnote{present address: Institut f\"ur Kernphysik,
  Universit\"at zu K\"oln, 550937 K\"oln, Germany }
  M.~B\"uscher,$^1$
  V.~Chernyshev,$^2$ 
  S.~Dymov,$^{1,3}$
  P.~Fedorets,$^2$ 
  V.~Grishina,$^4$ 
  C.~Hanhart,$^1$
  M.~Hartmann,$^1$ 
  V.~Hejny,$^1$
  A.~Khoukaz,$^5$ 
  H.R.~Koch,$^1$
  V.~Komarov,$^3$ 
  L.~Kondratyuk,$^2$ 
  V.~Koptev,$^6$
  N.~Lang,$^5$
  S.~Merzliakov,$^3$ 
  S.~Mikirtychiants,$^6$
  M.~Nekipelov,$^{1,6}$
  H.~Ohm,$^1$
  A.~Petrus,$^{3}$
  D.~Prasuhn,$^1$ 
  R.~Schleichert,$^1$ 
  A.~Sibirtsev,$^1$ 
  H.J.~Stein,$^1$ 
  H.~Str\"oher,$^1$ 
  K.--H.~Watzlawik,$^1$ 
  P.~W\"ustner,$^7$
  S.~Yaschenko,$^3$ 
  B.~Zalikhanov,$^3$ 
  I.~Zychor$^8$}

\affiliation{$^1$Institut f\"ur Kernphysik, Forschungszentrum J\"ulich, 52425 J\"ulich, Germany
}
\affiliation{$^2$Institute for Theoretical and Experimental Physics,
  Cheremushkinskaya 25, 117259 Moscow, Russia
}
\affiliation{$^3$Laboratory of Nuclear Problems, Joint Institute for
  Nuclear Research, 141980 Dubna, Russia
}
\affiliation{$^4$Institute for Nuclear Research, 60th October Anniversary
  Prospect 7A, 117312 Moscow, Russia
}
\affiliation{$^5$Institut f\"ur Kernphysik, Universit\"at M\"unster,
         W.--Klemm--Str.\ 9, 48149 M\"unster, Germany
}
\affiliation{$^6$High Energy Physics Department, Petersburg Nuclear
  Physics Institute, 188350 Gatchina, Russia
}
\affiliation{$^7$Zentralinstitut f\"ur Elektronik, Forschungszentrum
  J\"ulich, 52425 J\"ulich, Germany
} 
\affiliation{$^8$The Andrzej Soltan Institute for Nuclear Studies, 05400
  Swierk, Poland
}

\date{\today}

\begin{abstract}
  The reaction $pp {\to} d K^+ \bar{K^0}$ has been investigated at an
  excess energy of $Q{=46}$\,MeV above the $K^+\bar{K^0}$ threshold
  with ANKE at COSY--J\"ulich. From the detected coincident $dK^+$
  pairs about 1000 events with a missing $\bar{K^0}$ were identified,
  corresponding to a total cross section of $\sigma(pp {\to} d K^+
  \bar{K^0}) {=} (38 {\pm} 2_\mathrm{stat} {\pm}
  14_\mathrm{syst})$\,nb.  Invariant--mass and angular distributions
  have been jointly analyzed and reveal $s$--wave dominance between
  the two kaons, accompanied by a $p$--wave between the deuteron and
  the kaon system. This is interpreted in terms of
  $a_0^+(980)$--resonance production.
\end{abstract}
\pacs{13.75.Cs, 25.40.Ve}
\maketitle

One of the primary goals of hadronic physics is the description of the
internal structure of mesons and baryons, their production and decays,
in terms of quarks and gluons. However the severe complications,
related to nonperturbative contributions and confinement effects in
QCD, mean that progress in understanding the structure of light
hadrons has mainly been within models which use effective degrees of
freedom. The constituent quark model is one of the most successful in
this respect (see e.g.\ \cite{Morgan}).  This approach treats the
lightest scalar resonances $a_0/f_0(980)$ as conventional $\bar{q}q$
states.  However the structure of these states seems to be more
complicated and they have also been identified with $K\bar{K}$
molecules\,\cite{Weinstein} or compact $qq$--$\bar{q}\bar{q}$
states\,\cite{Achasov}.  It has even been suggested that at masses
below 1.0 GeV a complete nonet of 4--quark states might
exist\,\cite{Close}.  Possible deviations from the minimal quark model
is also generating much interest in the baryon sector, where the
recently discovered low-lying $\theta^+$ state in the $K^+n$
system~\cite{penta} requires at least five quarks.

At COSY--J\"ulich\,\cite{cosy} an experimental program has been
started, using the ANKE spectrometer\,\cite{ANKE_NIM}, which aims at
exclusive data on $a_0/f_0$ production from $pp$, $pn$,
$pd$ and $dd$ interactions close to the $K\bar{K}$
threshold\,\cite{css2002}.  The final goal will be the extraction of
the $a_0/f_0$ mixing amplitude to shed light on the nature of the
light scalar resonances. Here, we report about the first exclusive
study of the reaction $pp {\to} dK^+\bar{K^0}$, where some fraction of
the kaon pairs is expected to stem from the decay of the
$a_0^+$--resonance.  The measurements were performed at a beam energy
of $T_p{=}2.65$\,GeV ($p_p{=}3.46$\,GeV/c), corresponding to an excess
energy of 46\,MeV above the $K\bar{K}$ threshold.

The isovector $a_0(980)$ has been studied in $p\bar{p}$
annihilations~\cite{Amsler}, in $\pi^- p$ collisions~\cite{Teige}, and $\gamma
\gamma$ interactions~\cite{L3}.  Data on radiative
$\phi$--decays~\cite{SND,KLOE} are interpreted in terms of $a_0/f_0$
production in the channels $\phi {\to} \gamma a_0/f_0 {\to} \gamma \pi^0\eta/
\pi^0\pi^0$. In $pp$ reactions the $a_0(980)$ has been seen at
$T_p{=}450$\,GeV~\cite{WA102}, and a resonant structure around 980\,MeV/c$^2$
has been observed in inclusive $pp {\to} dX^+$ data at $p_p{=}3.8$, 4.5 and
6.3\,GeV/c~\cite{LBL}.

ANKE is located at an internal target position of COSY, which supplies
stored proton beams with intensities up to ${\sim} 4 {\times}
10^{10}$. A $H_2$ cluster--jet target\,\cite{target} has been used, providing areal
densities of ${\sim} 5 {\times} 10^{14}$\,cm$^{-2}$.  The luminosity
has been measured with the help of $pp$ elastic scattering events by
detecting one fast proton concurrently recorded with the $dK^+$
data\,\cite{QNP2002}.  Protons in the angular range $\vartheta{=}
5.5^\circ {-} 9^\circ$ have been selected, since the ANKE acceptance
changes smoothly for these angles and the elastic peak in the momentum
distribution is easily distinguished from the background.  The average
luminosity during the measurements has been determined to $L{=}(2.7
{\pm} 0.1_\mathrm{stat} {\pm} 0.7_\mathrm{syst}) {\times}
10^{31}$\,s$^{-1}$\,cm$^{-2}$ corresponding to
$L_\mathrm{int}{=}3.3$\,pb$^{-1}$.

ANKE comprises three dipole magnets (D1--3), which guide the
circulating COSY beam. The central C--shaped spectrometer dipole D2
downstream of the target separates the
reaction products from the beam. The angular acceptance of D2 for
kaons from $a_0^+$ decay is $|\vartheta_{\mathrm H}| {\leq} 12^{\circ}$
horizontally and $|\vartheta_{\mathrm V}| {\leq} 3.5^{\circ}$
vertically\,\cite{K_NIM}. The angular acceptance for the fast deuterons
($p_d{\sim} 2100$\,MeV/c) is roughly $|\vartheta_{\mathrm H}| {\leq}
10^{\circ}$ and $|\vartheta_{\mathrm V}| {\leq} 3.0^{\circ}$ and depends
on the momentum in the horizontal direction\,\cite{d_breakup}.

Positively charged kaons are identified in range telescopes, located
at the side of D2 along the focal surface, providing excellent
kaon--vs.--background discrimination. The kaons were identified by
means of time--of--flight (TOF) and energy--loss ($\Delta E$)
measurements.  Details of the procedure can be found in
Ref.\,\cite{K_NIM}. For our measurements, only some of the telescopes
were used, covering the lower momentum range
($p_{K^+}{=}(390 {-} 625)$\,MeV/c) of the $a_0^+$ decay $K^+$ mesons. Two
multi--wire proportional chambers (MWPCs) positioned in front of the
telescopes allow to deduce the ejectile momenta and to suppress
scattered background\,\cite{ANKE_NIM,K_NIM}.  Coincident fast particles
and elastically scattered protons are detected in the ANKE
forward--detection (FD) system consisting of two layers of
scintillation counters for TOF and $\Delta E$ measurements as well as
three MWPCs, each with two sensitive planes, for momentum
reconstruction and background suppression\,\cite{ANKE_NIM,d_breakup}.

\begin{figure}[h]
  \psfig{file=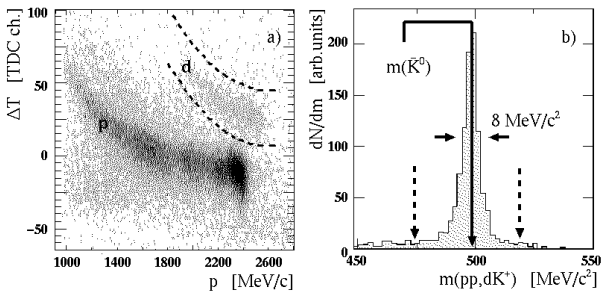,width=8.cm} 
  \vspace*{-4mm}
  \caption{a) Time difference between the fast 
    forward going particles in layer 1 of the FD scintillators and the
    $K^+$--mesons versus the momentum of the forward particle. The
    dashed lines indicate the selection for deuteron identification.
    b) Missing mass $m(pp,dK^+)$ distribution of the $pp {\to} dK^+X$ events.}
  \label{fig:tof_p}
\end{figure}

Two bands from protons and deuterons are clearly seen in the time
difference between the detection of a $K^+$--meson in one of the
telescopes and a particle in the FD as a function of the FD particle
momentum (see Fig.\,\ref{fig:tof_p}a). The deuterons are selected
with the criterion indicated by the dashed lines.  In
Fig.\,\ref{fig:tof_p}b) the missing--mass distribution $m(pp,dK^+)$
for the selected $pp {\to} d K^+ X$ events is presented. At
$T_p{=}2.65$\,GeV the missing particle $X$ must be a $\bar{K^0}$ due
to charge and strangeness conservation. The measured
$dK^+$ missing--mass distribution peaks at
$m_{\bar{K^0}}{=}498$\,MeV/c$^2$, reflecting the clean particle
identification. Approx. 1000 events within the gate indicated by the
dashed arrows are accepted as $dK^+\bar{K^0}$ events for further
analysis. The remaining background from misidentified particles is
($9 {\pm} 2$)\,\%.

The tracking efficiency for kaons in the side MWPCs has been
determined by requiring TOF and $\Delta E$ of a kaon (and any particle
in FD) and calculating the ratio of identified kaons (as peak in the
TOF distribution) with and without demanding a reconstructed
track. The efficiency depends on the telescope number, i.e.\
$K^+$--momentum, and varies between 71\,\% and 93\,\%. The efficiency
of the $\Delta E$ criterion for the $K^+$ mesons has been deduced with
the help of simultaneously recorded $pp {\to} pK^+\Lambda$ events
which --- due to the significantly larger cross section --- can be
selected by TOF alone. This efficiency is independent on the
telescope number and amounts to 52\,\%.  The FD MWPC efficiencies
have been determined for each of the six sensitive planes individually
from events with hits in all other five planes.  The information from
two horizontal (vertical) planes has been used to reconstruct the
intersection point in the remaining plane and, subsequently, to
determine the efficiency distribution across the chamber areas,
i.e.\ the angular and momentum dependences. The average FD track
efficiency for deuterons is 73\,\%.  The efficiency of the FD
scintillators and all TOF criteria is larger than
99\,\%\,\cite{K_NIM}. The data have been corrected for all
efficiencies on an event--by--event basis.

\begin{figure}[htb]
    \psfig{file=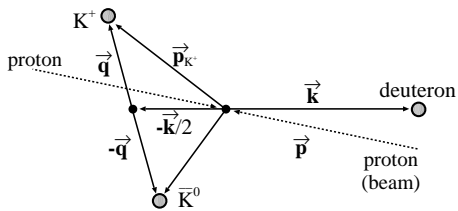,width=6cm} 
   \vspace*{-4mm}
    \caption{Definition of the vectors $\vec{p}$, $\vec{k}$ and 
      $\vec{q}$ in the cms of the reaction $pp {\to}
      dK^+\bar{K^0}$. Angular distributions with respect to the beam
      direction $\vec{p}$ have to be symmetric around $90^\circ$ since
      the two protons in the entrance channel are indistinguishable.}
    \label{fig:coordinates}
\end{figure}

\begin{figure}[hb]
  \psfig{file=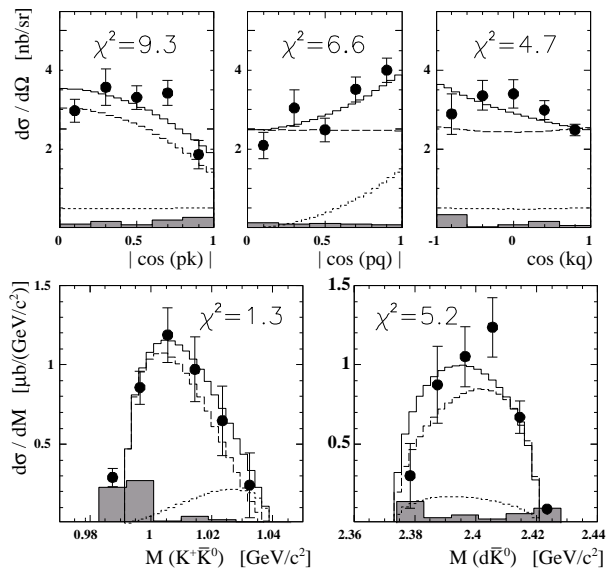,width=8.cm}
  \vspace*{-4mm} 
  \caption{Angular and invariant mass distributions.  The
  shaded areas correspond to the systematic uncertainties of the
  acceptance correction. The dashed (dotted) line corresponds to
  $K^+\bar{K^0}$ production in a relative $s$-- ($p$--) wave and the solid
  line is the sum of both contributions. The overall errors from the
  luminosity determination are not included in the statistical and
  systematic uncertainties. The common fit with Eq.\ref{eq:dcs} to all
  spectra yields $\chi^2_{\mathrm{ndf}}=1.1$. The individual
  $\chi^2$ values are displayed in the corresponding panels.}
  \label{fig:differential_spectra}
\end{figure}

The differential acceptance of ANKE has been obtained with the
Monte--Carlo method described in Ref.\,\cite{disto}, which allows
one to determine the acceptance independently of the ejectile
distributions at the production point. The acceptance is defined as a
discrete function of the five relevant degrees of freedom in the
3--body final state. For an unpolarized measurement the acceptance can
be expressed by a 4--dimensional matrix with four independent
kinematical variables.  Two different matrices, each composed of 500
elements (see Table~\ref{tab:elem}), were used for the reconstruction
of the invariant masses $m(K^+\bar{K^0})$, $m(d\bar{K^0})$ and the
center--of--mass (cms) angular distributions $|\cos(pk)|$,
$|\cos(pq)|$, $\cos(kq)$ (for a definition see
Fig.\,\ref{fig:coordinates}).  90 million events following a
phase--space distribution were simulated and tracked through ANKE,
taking into account small angle scattering, decay in flight and the
MWPC resolutions. Subsequently, the missing mass and angular
distributions have been corrected with the weights from the
corresponding acceptance matrices on an event--by--event
basis. The impact of the few acceptance holes has been
investigated using distributions with different shapes in the masses
and angles and is included in the systematic error of the differential
distributions and the total cross section. The efficiency and
acceptance corrected data are shown in
Fig.\,\ref{fig:differential_spectra} with statistical error bars and
systematic uncertainties.

The integrated acceptance of ANKE for the $dK^+\bar{K^0}$ events is
$2.2$\,\%.  Taking into account all correction factors a total cross
section of $\sigma(pp {\to} d K^+ \bar{K^0}) {=} (38
{\pm} 2_\mathrm{stat} {\pm} 14_\mathrm{syst})$\,nb has been derived.

\begin{table}[htb]
  \caption{Variables (all in the overall cms) of acceptance matrices I
    (upper) and II (lower) and their number of bins. Matrices I and II
    are used for the correction of the mass and angular spectra,
    respectively. $\psi(K\bar{K})$ is the rotation angle of the decay
    plane around the direction of the deuteron momentum $\vec{k}$. The
    symmetry of the angular distributions around $90^\circ$ with
    respect to the proton beam has been utilized.}
    \begin{ruledtabular} \begin{tabular}{cccc} \label{tab:elem}
    $m^2(K^+\bar{K^0})$ & $m^2(d\bar{K^0})$ & $|\cos(pk)|$ &
    $\psi(K\bar{K})$\\ 5 & 5 & 5 & 4\\ \colrule $|\cos(pk)|$ &
    $|\cos(pq)|$ & $\cos(kq)$ & $E(d)$\\ 5 & 5 & 5 & 4\\ \end{tabular}
    \end{ruledtabular}
\end{table}

The mass and angular resolutions for the spectra in
Fig.\,\ref{fig:differential_spectra} have been obtained from
simulations and amount to: $\delta
m_{K^+\bar{K^0}}{=}(8{-}1)$\,MeV/c$^2$ in the range $(0.991{-}
1.038)$\,GeV/c$^2$, $\delta m_{d\bar{K^0}} {\sim} 3$\,MeV/c$^2$ in the
full mass range, and $\delta(\cos{(\vartheta)}) {\sim} 0.2$ for all
angular spectra (FWHM values).  The effect of these resolutions on the
shape of the differential spectra is included in the systematic errors
shown in Fig.\ \ref{fig:differential_spectra}.  Due to the finite
values of $\delta m$ some events lie outside the kinematical limits at
$m(K^+\bar{K^0}){=}0.991$ and $m(d\bar{K^0}){=}2.420$\,GeV/c$^2$,
respectively.  These data points have been excluded from the fit
described below.

In the close--to--threshold regime only a limited number of final states
can contribute. If we restrict ourselves to the lowest partial waves,
we need to consider either a $p$--wave in the $K \bar K$ system
accompanied by an $s$--wave of the deuteron with respect to the meson
pair or an $s$--wave $K \bar K$ with a $p$--wave deuteron. Selection
rules do not allow for $s$--waves in both subsystems simultaneously.
Under these assumptions one may write for the square of the transition
matrix element
\begin{eqnarray}
  \nonumber \bar{|{\cal M}|^2}&=& C_0^qq^2+C_0^kk^2+C_1(\hat p
  \cdot \vec k)^2+C_2(\hat p \cdot \vec q)^2
  \\
  & & \ \ \ \ \ \ \ +C_3(\vec k \cdot \vec q)+C_4(\hat p \cdot
  \vec k)(\hat p \cdot \vec q) \ ,
  \label{eq:dcs}
\end{eqnarray}
with $\hat p {=}\vec p/|\vec p|$. The parameters $C_i$ can be directly
related to the 8 allowed partial wave amplitudes, as was done e.g.\ in
Refs.\,\cite{Grishina_PLB,Tarasov,kthhs}.  Only $K\bar{K}$ $p$--waves
contribute to $C_0^q$ and $C_2$, only $K\bar{K}$ $s$--waves to $C_0^k$
and $C_1$, and only $s$-$p$ interference terms to $C_3$ and $C_4$.
Thus, a fit to the data supplies direct information on the relative
strength of the different partial waves.  The result of the fit is
shown as the solid line in Fig.\,\ref{fig:differential_spectra} and
the corresponding parameters $C_i$ are displayed in Table
\ref{tab:fit} (note that from our spectra $C_3$ and $C_4$ cannot be
extracted individually but only ($C_3+(1/3)C_4$)). The overall
agreement between fit and data confirms the assumption that only the
lowest partial waves contribute. The dashed lines in
Fig.\,\ref{fig:differential_spectra} depict the contributions from
$K\bar{K}$ $s$--waves only ($C_0^q{=}C_2{=}C_3{=}C_4{=}0$) and the dotted line
those from pure $K\bar{K}$ $p$--waves ($C_0^k{=}C_1{=}C_3{=}C_4{=}0$). Thus,
we find that the reaction is dominated by the $K\bar{K}$ $s$--wave.
The contribution from $K\bar{K}$ $p$--waves is less than 20\%.

\begin{table}[htb]
  \caption{Result for the various coefficients of Eq.\,\ref{eq:dcs}. All
    coefficients are given in units of GeV$^{-2}$.}
  \begin{ruledtabular}
    \begin{tabular}{ccccc}
      \label{tab:fit}
      $C_0^q$ & $C_0^k$ & $C_1$ & $C_2$ & $C_3+\frac{1}{3}C_4$\\ \colrule    
      $0{\pm}0.1$ & $0.89{\pm}0.03$ & $-0.5{\pm}0.1$ & $1.12{\pm}0.07$ & $-0.32{\pm}0.15$\\
    \end{tabular}
  \end{ruledtabular}
\end{table}

Since the $K \bar{K}$ $p$--waves prefer large invariant masses, it is
the slope at higher $K \bar{K}$ masses that sets an upper bound for
the $p$--wave contribution.  On the other hand, the slope of the
$|\cos (pq)|$ distribution determines the minimum $p$--wave strength.
Therefore, the contribution of $K \bar{K}$ $p$--waves is strongly
constrained by the data.  The deviation in the upper left panel might
indicate a small contribution from a $d$--wave of the deuteron with
respect to the $s$--wave meson pair. Note that this $d$--wave contribution
would have only little influence on the shapes of the other measured
distributions.

Since the excess energy is small and the nominal mass of the
$a_0^+$ is very close to the $K \bar{K}$ threshold, the resonance
cannot be seen as a clear structure in the invariant mass spectra but
in any case the propagation of low energy $s$--wave $K\bar{K}$ pairs
should be governed by the $a_0^+$--resonance due to the proximity of
its pole to the $K\bar{K}$ threshold. This, together with the result
of the fit leads us to the conclusion that the reaction $pp {\to}
dK^+\bar{K}^0$ at $T_p{=}2.65$\,GeV proceeds mainly via the
$a_0^+$--resonance.

\begin{figure}[tb]
    \psfig{file=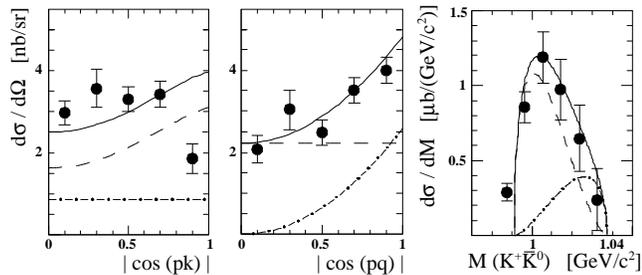,width=8.5cm} 
    \vspace*{-4mm}
    \caption{Predictions of the model of Ref.\
    \protect{\cite{Grishina_MESON2002}} in comparison to the
    experimental data.  An overall scaling factor is included as
    described in the text.}  \label{fig:model}
\end{figure}

In Ref.\,\cite{Grishina_MESON2002} a model has been presented which
describes both resonant, dominated by the direct $a_0$ production off
a single nucleon, as well as non--resonant $K \bar{K}$ production,
driven by a $\pi{-}K^*{-}\pi$ exchange current.  This current, due to
G--parity conservation, generates the $K\bar{K}$ pair in a relative
$p$--wave. The only free parameter of the model, an overall factor,
was adjusted to higher energy data\,\cite{LBL}.  The predictions of
this model for some of the spectra measured in this experiment are
shown in Fig.\,\ref{fig:model}.  A scaling factor of 0.75 has been
included in order to adjust the model results to the integrated cross
section.

As can be seen from the figure the model calculations agree with the
data for $m(K^+\bar{K}^0)$ and $|\cos(pq)|$, the spectra which are
most sensitive to the relative $s$-- and $p$--wave contributions.
The discrepancy in Fig.\,\ref{fig:model}, left side,
indicates that there might be another possible reaction mechanism
involved.  Here it is important to note that within the model from
Ref.\,\cite{Grishina_MESON2002} the ratio of integrated resonant to
non--resonant production cross section has very little sensitivity on
the details of the $\cos(pk)$ distribution.

To summarize, first data on the reaction $pp {\to} d K^+\bar{K^0}$ close
to the production threshold are presented. The total cross section for
this reaction as well as differential distributions have been
determined.  Dominance of the $K\bar{K}$ $s$--wave ($>$~80\,\% of
the total cross section) has been deduced based on a joint analysis of
angular and invariant mass spectra. This is a clear evidence for a
predominant resonant production via the $a_0^+(980)$.  A comparison of
the experimental data to model calculations\,\cite{Grishina_MESON2002}
shows, that the background ($K \bar{K}$ $p$--waves) can be understood
in terms of a simple meson exchange current. Further theoretical
investigations are necessary to identify the possible role of a strong
$d\bar{K}$ interaction as proposed in Ref.\,\cite{a0_Oset}.  Here both
a measurement at higher energies and polarization experiments would be
of great use.
Our results demonstrate the feasibility of
studying light scalar resonances in exclusive measurements close to
the $K \bar{K}$ threshold with high mass resolution and low
background using ANKE at COSY-J\"ulich.

We regret to announce that one of the authors, Sasha Petrus, died in a
tragic accident on May 19, 2002.  We are grateful to W.~Borgs,
C.~Schneider and the IKP technicians for the support during the
beamtime and to J.~Ritman and C.~Wilkin for stimulating discussions
and careful reading of the manuscript.  The work at ANKE has partially
been supported by: BMBF (grants WTZ-RUS-684-99, 686-99, 211-00,
691-01, GEO-001-99, POL-007-99, 015-01), DFG (436 RUS 113/444, 561,
630); Polish State Committee for Scientific Research (2 P03B 101 19);
Russian Academy of Science (RFBR99-02-04034, 18179a, 06518,
02-02-16349); ISTC (1861, 1966).

\end{document}